\documentclass[runningheads,a4paper]{llncs}

\usepackage{amssymb}
\usepackage{amsmath}

\usepackage{graphicx}
\usepackage{comment}

\usepackage{algorithmic}
\usepackage{algorithm} 
\usepackage{setspace}

\usepackage{listing}
\usepackage{listings}
\usepackage{tabularx}

\usepackage{authblk}
\usepackage{hyperref}
\usepackage{subcaption}
\usepackage{fancyvrb}
\usepackage{float}

\usepackage{xcolor}
\definecolor{mygreen}{RGB}{28,172,0} 
\definecolor{mylilas}{RGB}{170,55,241}
\definecolor{myazure}{RGB}{0,175,225}
\definecolor{myorange}{RGB}{175,125,0}
\definecolor{myart}{RGB}{255,0,0}
\definecolor{review}{RGB}{255,0,0}
\definecolor{myrev}{RGB}{0,0,0}

\definecolor{myart}{RGB}{0,0,0}
\definecolor{review}{RGB}{0,0,0}
\definecolor{myrev}{RGB}{0,0,0}

\usepackage{url}
\usepackage{hyperref}

\usepackage{tikz}
\usetikzlibrary{external}
\usepackage{tabularx}
\usepackage{tabulary}

\urldef{\mailjv}\path|jan.valdman@utia.cas.cz |
\urldef{\mailam}\path|alexmos@kma.zcu.cz |
\urldef{\mailmf}\path|mfrost@it.cas.cz |

\newcommand{\nt}{|\mathcal{T}|}
\newcommand{\nn}{|\mathcal{N}|}
\newcommand{\ned}{|\mathcal{E}|}

\newcommand{\R}{\mathbb R}
\newcommand{\N}{\mathbb N}

\newcommand{\dxy}{\, \mathrm{d}\boldsymbol{\mathrm{x}}}

\newcommand{\m}{m}
\newcommand{\p}{p}
\renewcommand{\P}{p}
\newcommand{\s}{s}
\newcommand{\nb}{n_{\p}}
\newcommand{\nip}{n_{ip}}
\newcommand{\Tr}{T_{ref}}
\newcommand{\power}{\alpha}

\newcommand{\nbref}{n_{\p,ref}}
\newcommand{\SpT}{S^{\P}(\mathcal{T})}

\newcommand{\x}{\textbf{x}}
\newcommand{\f}{\textbf{f}}
\newcommand{\vv}{\textbf{v}}
\newcommand{\fx}{\f(\x)}
\newcommand{\vx}{\vv(\x)}
\newcommand{\F}{\textbf{F}}

\newcommand{\textbfn}[1]{'\textbf{#1}'}

\begin{document}

\mainmatter  
\title{Minimization of energy functionals via FEM: implementation of hp-FEM}

\titlerunning{Minimization of energy functionals via FEM}

\author{Miroslav Frost\inst{1}
\and
Alexej Moskovka\inst{2}
\and
Jan Valdman\inst{3,4}
\thanks{
A. Moskovka and J. Valdman announce the support of the Czech Science Foundation (GACR) through the grant 21-06569K. M. Frost acknowledges the support of the Czech Science Foundation (GACR) through the grant 22-20181S.}
}

\authorrunning{Miroslav Frost, Alexej Moskovka, Jan Valdman}

\institute{$^1$Institute of Thermomechanics, Czech Academy of Sciences, \\
Dolej\v{s}kova 5, 18200 Prague, Czech Republic \\
\mailmf \\
\vspace{0.15cm}
$^2$Department of Mathematics, Faculty of Applied Sciences, \\
University of West Bohemia, 
Technick\' a 8, 30100 Pilsen, Czech Republic \\
\mailam \\
\vspace{0.15cm}
$^3$The Czech Academy of Sciences, Institute of Information Theory \\ and Automation, 
Pod Vod\'{a}renskou v\v{e}\v{z}\'{\i}~4, 18208~Prague, Czech Republic  \\
$^4$Faculty of Information Technology, Czech Technical University in Prague, Th\'{a}kurova 9, 16000 Prague, Czech Republic \\
\mailjv
}
\maketitle
\begin{abstract}
Many problems in science and engineering can be rigorously recast into minimizing a suitable energy functional. We have been developing efficient and flexible solution strategies to tackle various minimization problems by employing finite element discretization with P1 triangular elements \cite{MMV,MoVa}. An extension to rectangular hp-finite elements in 2D is introduced in this contribution.

\end{abstract}

\section{Introduction}
The finite element method (FEM) can be efficiently used to minimize energy functionals appearing in various types of problems.
The simplest P1 finite elements were implemented in MATLAB for the discretization of p-Laplace energy functional in \cite{MMV}. We introduced several vectorization techniques in \cite{MoVa} for an efficient evaluation of the discrete energy gradient  and, additionally, applied these techniques for the minimization of hyperelasticity in 2D and 3D. Recently, our approach has been successfully applied to 2D/3D problems in solid mechanics, namely the resolution of elastoplastic deformations of layered structures \cite{KSV} or superelastic and pseudoplastic deformations of shape-memory alloys \cite{FV}. 

The hp-FEM is an advanced numerical method based on FEM dating back to the pioneering works of I. Babuška, B. A. Szabó and co-workers in 1980s, e.g. \cite{BabuskaSzabo1} and \cite{BabuskaSzabo2}. It provides increased flexibility and convergence properties compared to the ``conventional" FEM. {\color{myrev}There are recent MATLAB implementations including triangular elements \cite{MOOAFEM} and rectangular elements \cite{HP}}.

In this {\color{myrev}paper},
we combine energy evaluation techniques of \cite{MoVa} and the hp-FEM implementation \cite{HP}. The trust-region (TR) method \cite{conn2000} is applied for the actual minimization of energies. It is available in the MATLAB Optimization Toolbox and was found to be very efficient in the comparison performed in \cite{MoVa}.
It requires the gradient of a discrete energy functional and also allows to specify a sparsity pattern of the corresponding Hessian matrix which is directly given by a finite element discretization. We employ two different options:
\begin{itemize}
    \item[$\bullet$] option 1: the TR method with the gradient evaluated directly via its explicit form and the specified Hessian sparsity pattern.
    \item[$\bullet$] option 2: the TR method with the gradient evaluated approximately via central differences and the specified Hessian sparsity pattern.
\end{itemize}

 We demonstrate the capabilities of our implementation on two particular problems in 2D: the scalar p-Laplace problem and the vector hyperelasticity problem. The underlying MATLAB code
  is available at
\begin{center}
\url{https://www.mathworks.com/matlabcentral/fileexchange/125465} 
\end{center}
for free download and testing. Running times were obtained on
a MacBook Pro (M2 Pro processor, 2023) with 16 GB memory running MATLAB R2023a.

\section{Hierarchical shape basis functions}
Given a reference element $\Tr = [-1,1]^2$ and $\P \in \N$ we define by $S^{\P}(\Tr)$ a space of all (local) shape basis functions of polynomial degrees less or equal {\color{myrev}to} $\p$ defined on $\Tr$ and we denote by $\nbref$ their number. The construction of these functions is based on Legendre polynomials and is described in detail in \cite{BabuskaSzabo2}.
\begin{figure}[H]
    \vspace{-0.0cm}
    \centering
    \includegraphics[width=0.99\textwidth]{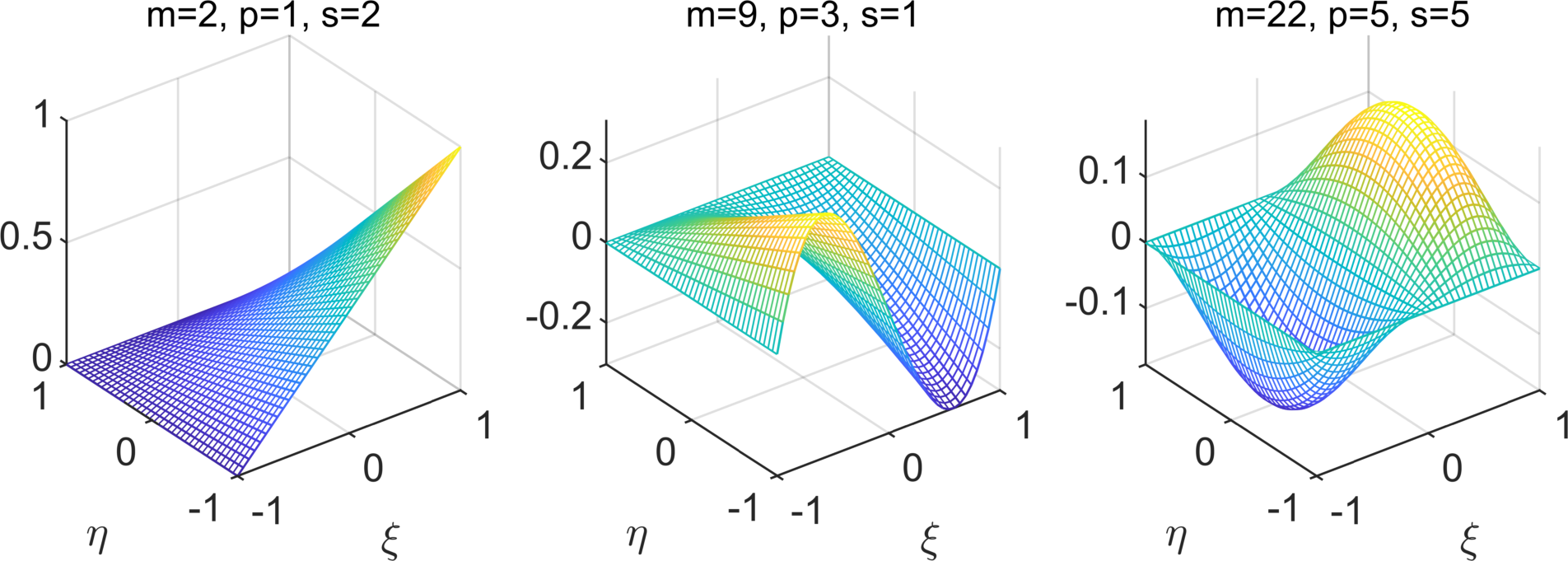}
    \caption{Example of nodal, edge, and bubble shape basis functions on $\Tr$. Generated by a modification of the `Benchmark 2' from \cite{HP}.}
    \vspace{-0.5cm}
    \label{fig:hp_basis_2D}
\end{figure}
Generally, there are three types of shape basis functions shown in Fig. \ref{fig:hp_basis_2D}:
\begin{itemize}
    \item the nodal (Q1) shape basis function of the first order is nonzero in one particular node and vanishes in all other nodes;
    \item the edge shape basis function of the $p$-th order is nonzero on one particular edge and vanishes on all other edges;
    \item the bubble shape basis function of the $p$-th order vanishes on the whole boundary of $\Tr$.
\end{itemize}

All shape basis functions on $\Tr$ are sorted by the polynomial degree $\p$ and the type of shape function. {\color{myrev}Every local basis function is assigned a unique index $\m$. It is determined by the polynomial order $\p$ and an additional index $\s$ given by the local index of a node, edge or bubble}.

We assume a computational domain $\Omega$ and its decomposition $\mathcal{T}$ into quadrilaterals in the sense of Ciarlet \cite{Ciarlet-FEM}. We denote by $\nn$, $\ned$ and $\nt$ the number of nodes, edges and elements, respectively.
The local shape basis functions are used for the construction of global ones defined on the whole $\mathcal{T}$. We denote by $\SpT$ the space of all global basis functions of polynomial degrees less or equal {\color{myrev}to} $\p$ and by $\nb$ their number. In \cite{HP} we introduced several key matrices providing the relation between the topology of $\mathcal{T}$ and the corresponding global basis functions. The first matrix collects the indices of all global basis functions and their type, the second collects for every element the indices of global basis functions that are nonzero on that element, and the third stores for every element the signs of local basis functions (necessary for edge functions of an odd degree).

\begin{example}
Quadrilaterals of the L-shape domain are shown in Fig. \ref{fig:hess_sparsity} (left) in which $\nn = 21$, $\ned = 32$, $\nt = 12$. For $\P = 2$ we have $\nb = 53$ global basis functions ($21$ nodal and $32$ edge) and the Hessian sparsity pattern (right) can be extracted directly.
\begin{figure}
\vspace{-0.4cm}
\centering
\begin{minipage}[c]{0.48\textwidth}
\includegraphics[width=0.99\textwidth]{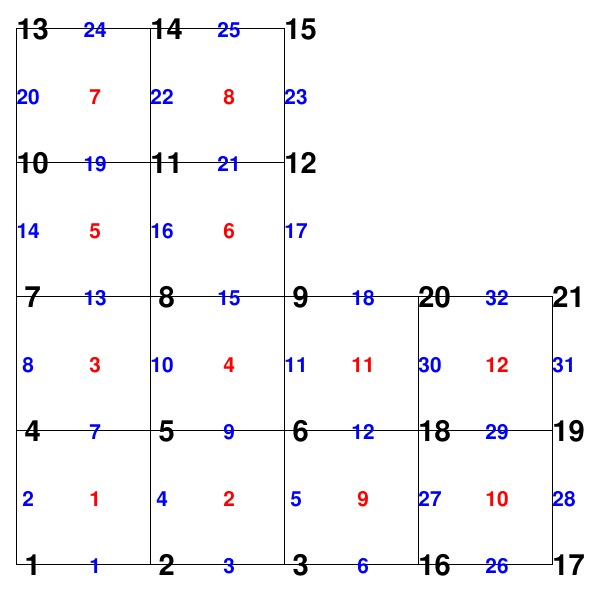}
\end{minipage}
\begin{minipage}[c]{0.48\textwidth}
\includegraphics[width=0.99\textwidth]{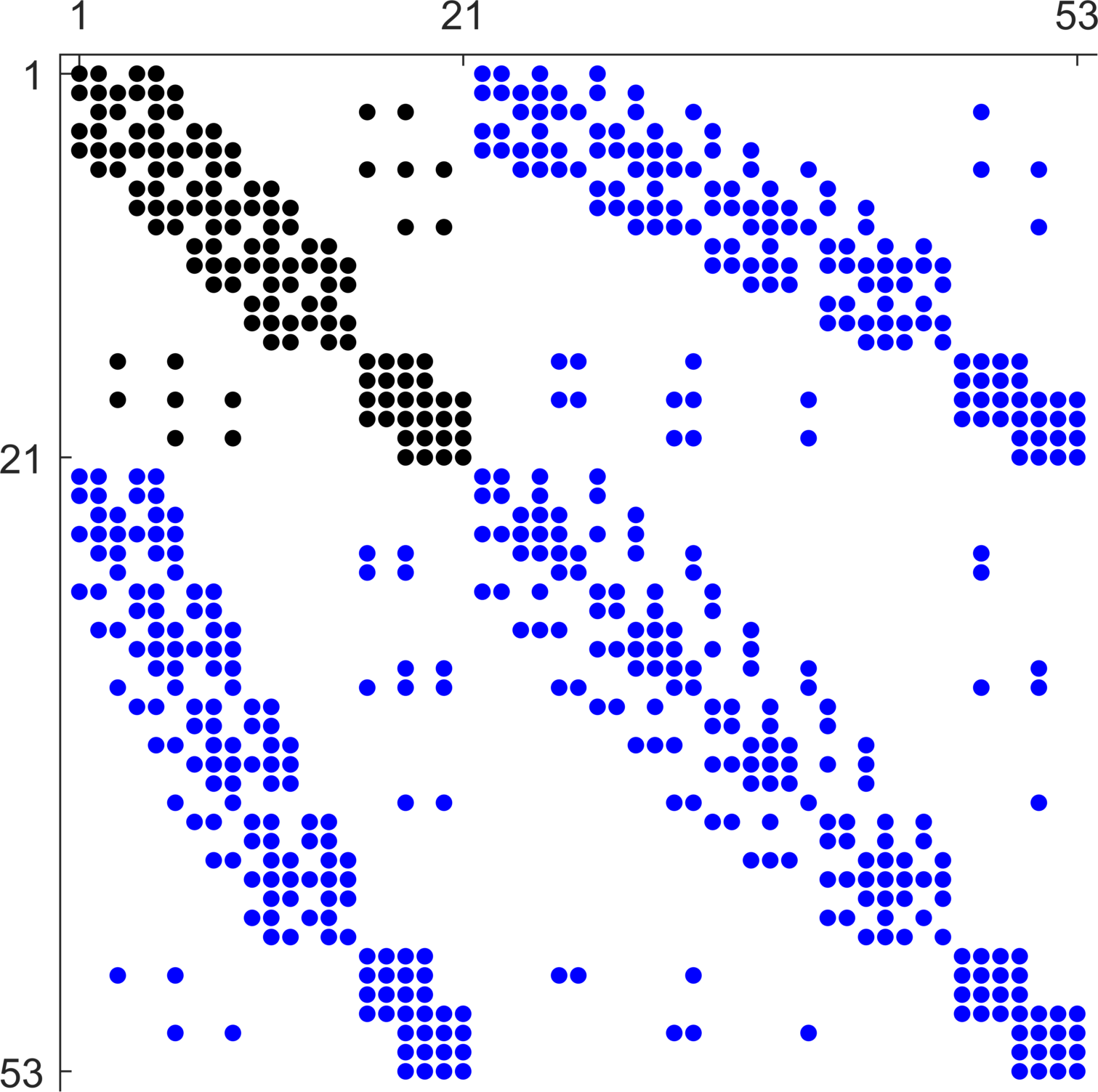}
\end{minipage}
\caption{A rectangular mesh (left) and the corresponding Hessian sparsity pattern for $\p = 2$ hierarchical elements (right). The {\color{myrev} upper diagonal submatrix} highlighted by the black color corresponds to Hessian sparsity for $\p = 1$.}
\label{fig:hess_sparsity}
\vspace{-0.6cm}
\end{figure}
\end{example}

\section{Models and implementation}

\subsection{p-Laplace equation} \label{subsec:pLaplace}

\begin{figure}[H]
\centering
\begin{minipage}[c]{0.48\textwidth}
\includegraphics[width=\textwidth]{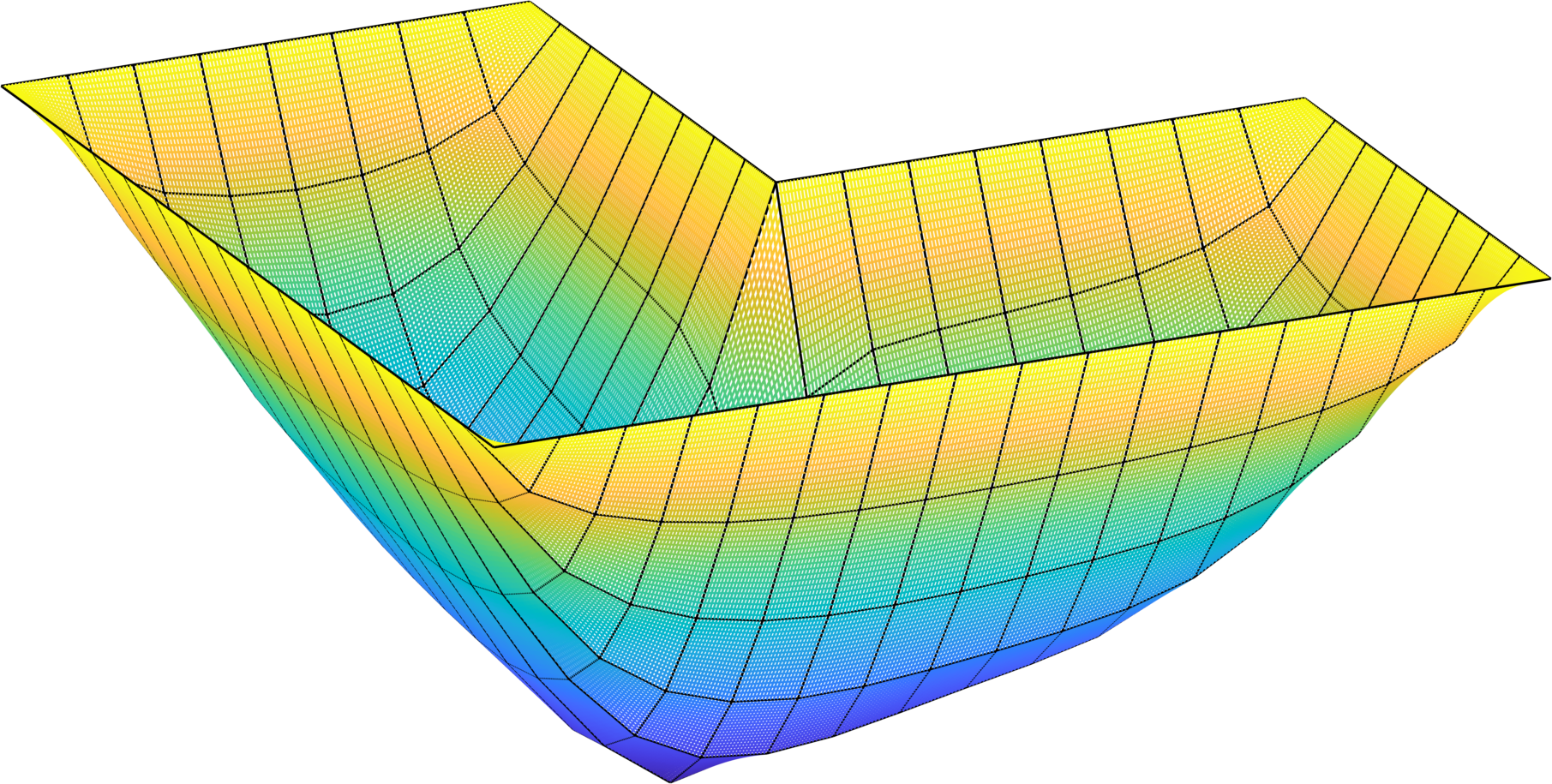}
\end{minipage} \:\:\:
\begin{minipage}[c]{0.48\textwidth}
\includegraphics[width=\textwidth]{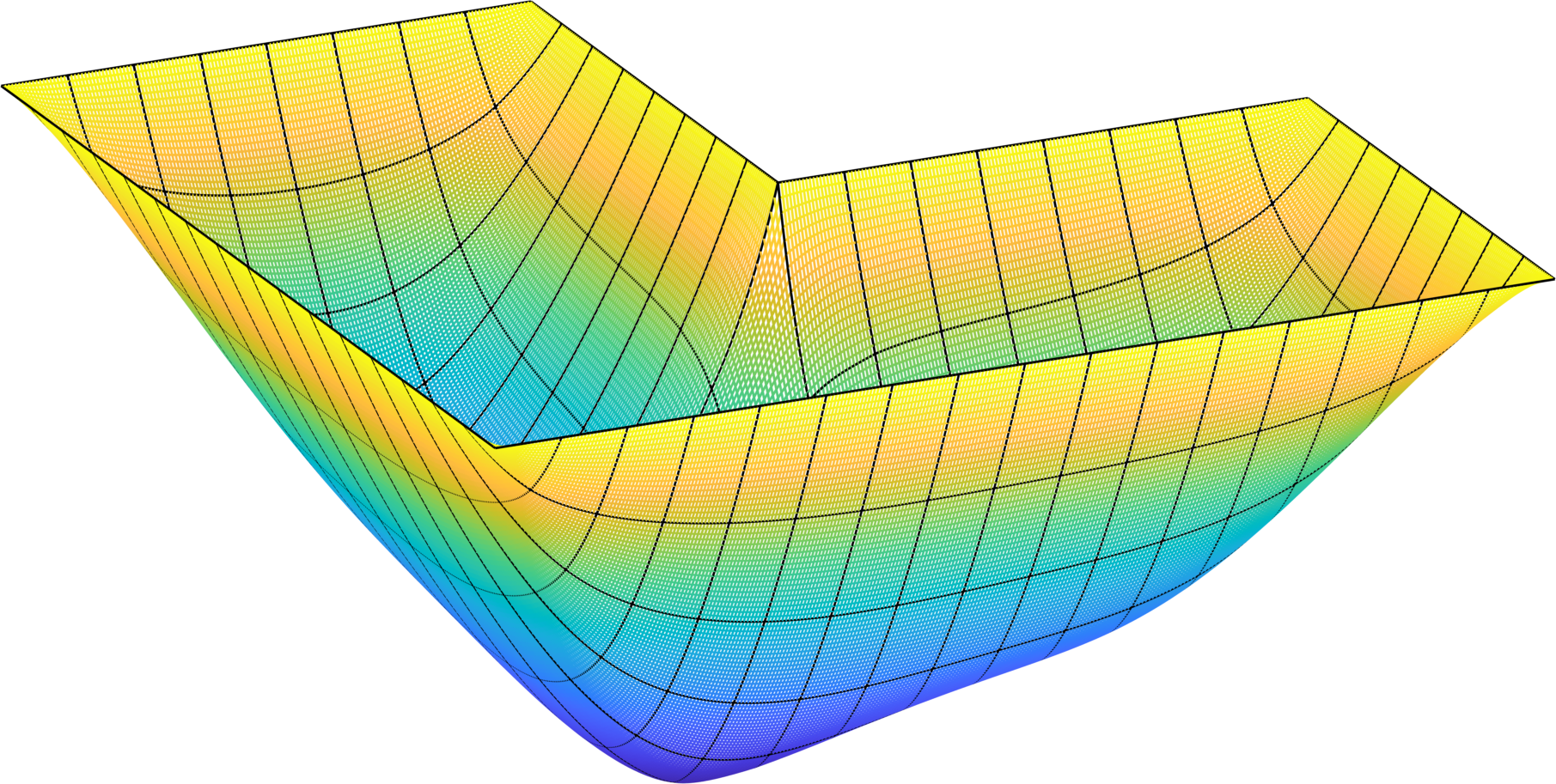}
\end{minipage}
\caption{Numerical solutions of the p-Laplacian with $\power = 3$ for a computational mesh with {\color{myrev}$\nt = 192$} and polynomial bases for $p = 1$ (left) and $p = 4$ (right).}
\label{pLaplace_2D}
\vspace{-0.6cm}
\end{figure}

\begin{figure}
    \vspace{-0.7cm}
    \centering
    \includegraphics[width=0.94\textwidth]{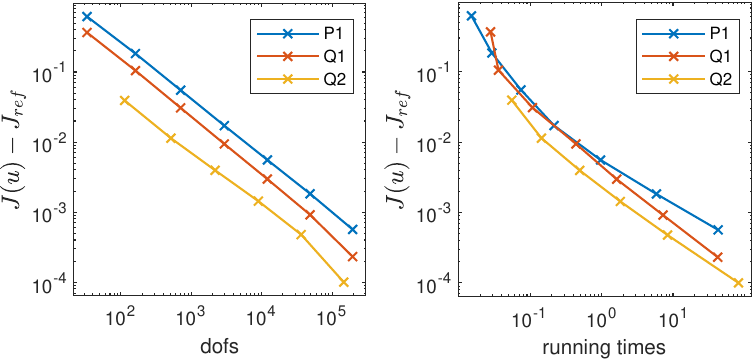}
    \vspace{-1mm}
    \caption{Performance of p-Laplacian for ${\power} = 3$: comparison of elements.}
    \label{fig:comparison_pLaplace}
    \vspace{-1.1cm}
\end{figure}

We are interested in a (weak) solution of the p-Laplace equation \cite{Lindqvist}: 
\begin{equation} \label{pLapl}
\begin{split}
\Delta_{\power} u &= f \qquad\quad \mbox{in} \:\: \Omega \, , \\
u &= g \qquad \,\, \mbox{on} \:\: \partial \Omega \, ,
\end{split}
\end{equation}
where the p-Laplace operator is defined as $
\Delta_{\power} u = \nabla \cdot \big( \|\nabla u\|^{{\power}-2} \nabla u \big)$ for some power ${\power}>1$ {\color{myrev}(the integer $p$ denotes the polynomial degree of $S^{\p}(\Tr)$)}. The domain $\Omega \in \mathbb{R}^d$ is assumed to have a Lipschitz boundary $\partial \Omega$, $f \in L^2(\Omega)$ and $g \in W^{1-1/{\power},{\power}}(\partial \Omega)$, where $L$ and $W$ denote standard Lebesque and Sobolev spaces. 
It is known that \eqref{pLapl} represents an Euler-Lagrange equation corresponding to a minimization problem
\begin{equation} \label{energy1D}
J(u)=\min_{v \in V}J(v) \, , \qquad J(v):=\frac{1}{{\power}} \int_{\Omega} \|\nabla v\|^{\power} \dxy - \int_{\Omega}  f \, v  \dxy \, ,
\end{equation}
where $V=W^{1,{\power}}_g(\Omega)=\{v \in W^{1,{\power}}, v = g \mbox{ on } \partial \Omega \}$ includes Dirichlet boundary conditions on $\partial \Omega$. The minimizer $u \in V$ of \eqref{energy1D} is known to be unique for ${\power} > 1$. {\color{myrev} It corresponds to the classical Laplace operator for ${\power} = 2$. The analytical handling of \eqref{pLapl} is difficult for general $f$}.  
The equation \eqref{pLapl} in 2D ($d=2$) takes the form
\begin{equation}
\nabla \cdot \Big( \big((\partial_1 u)^2 + (\partial_2 u)^2 \big)^{\frac{{\power}-2}{2}} \nabla u \Big) = f \qquad \mbox{in} \:\: \Omega
\end{equation}
and the corresponding energy reads
\begin{equation} 
J(v):=\frac{1}{{\power}} \int_{\Omega} \big((\partial_1 v)^2 + (\partial_2 v)^2 \big)^{\frac{{\power}}{2}} \dxy - \int_{\Omega}  f \, v  \dxy \, .
\end{equation}
{\color{myrev} The evaluation of integrals above is based on the Gaussian quadrature. We apply the number of Gauss points corresponding to the quadrature of order $p+1$. This does not guarantee the exact quadrature, but it proved to be sufficient in our numerical tests. 
} 
Figure \ref{pLaplace_2D} illustrates numerical solutions for the L-shape domain from Figure \ref{fig:hess_sparsity}, for a constant $f = -10$, ${\power} = 3$ and zero Dirichlet boundary condition on $\partial \Omega$. 
\noindent
Table~\ref{tab:pLaplace_Q2} shows the performance for 
Q2 elements (corresponding to the choice $p=2$). 
{\color{myrev} 
We notice that computations using the explicitly evaluated gradient are faster than using the numerical gradient (via central differences). 
A comparison to  Q1 elements (corresponding to the choice $p=1$) or P1 (triangular) elements of \cite{MoVa} is depicted in Fig. \ref{fig:comparison_pLaplace}. We observe a lower number of needed degrees of freedom (dofs) for Q2 elements and slightly lower running times to achieve the same accuracy. Since the exact energy is not known in this example, we use $J_{ref}$ as the smallest of all achieved energy values $J(u)$ of Table ~\ref{tab:pLaplace_Q2} decreased by $10^{-4}$.}
\vspace{-0.6cm}

\newcolumntype{d}{>{\hsize=1\hsize}X}
\newcolumntype{s}{>{\hsize=.5\hsize}X}
\newcolumntype{Y}{>{\raggedleft\arraybackslash}X}
\newcolumntype{Z}{>{\centering\arraybackslash}X}
\newcolumntype{S}{>{\centering\arraybackslash}s}

\begin{table}
    \centering
    \begin{tabularx}{0.99\textwidth}
    {S Y Y | Y  Z Y | Y Z Y}
      & & & \multicolumn{3}{c|}{explicit gradient} & \multicolumn{3}{c}{numerical gradient}  \\
      \hline
     level & $\nt$ & dofs  &  time [s] & iters & $J(u)$ & time [s] & iters & $J(u)$ \\
 \hline
1 & 48 & 113 &      0.06 & 7 &   -7.9209 &      0.07 & 7 &   -7.9209 \\ 
2 & 192 & 513 &      0.14 & 8 &   -7.9488 &      0.18 & 8 &   -7.9488 \\ 
3 & 768 & 2177 &      0.49 & 10 &   -7.9562 &      0.67 & 10 &   -7.9562 \\ 
4 & 3072 & 8961 &      1.73 & 12 &   -7.9587 &      2.38 & 12 &   -7.9587 \\ 
5 & 12288 & 36353 &      8.31 & 13 &   -7.9596 &     10.60 & 13 &   -7.9596 \\ 
6 & 49152 & 146433 &     80.81 & 13 &   -7.9600 &    136.92 & 14 &   -7.9600 \\ 
    \end{tabularx}
    \vspace{0.25cm}
 \caption{Performance of p-Laplacian for ${\power} = 3$ and Q2 elements.}
 \label{tab:pLaplace_Q2}
     \vspace{-0.7cm}
\end{table}

\subsection{Hyperelasticity}

Boundary value problems in (non-linear) elastostatics provide examples of vector problem which can be directly dealt with our approach, see \cite{MoVa}. Deformation, $\vx$, of a (hyper)elastic body spanning the domain $\Omega \in \R^{d}$ subjected to volumetric force, $\fx$, can be obtained by minimization of the corresponding energy functional, $J$, which takes the form:
\begin{equation}
   J(\vx) = \int_{\Omega} W\big(\F(\vx)\big) \dxy - \int_{\Omega} \fx \cdot \vx \dxy \,,
\end{equation}
where $\F(\vx) = \nabla \vx$ is deformation gradient and

\begin{equation} \label{neoHook}
    W(\F) = C_1 \big(I_1(\F)-\dim -2  \log(\det \F)\big) + D_1 (\det \F -1)^2 \, ,
\end{equation}
is so-called compressible Neo-Hookean energy density with $C_1, D_1$ being material constants {\color{myrev}and $I_1(\F) = \|\F\|^2$ denotes the squared Frobenius norm};
see \cite{MarsdenHughes} for details on the underlying continuum mechanics theory and its mathematically rigorous formulation. 
\begin{figure}[H]
\vspace{-0.5cm}
\centering
\begin{minipage}[c]{0.445\textwidth}
\includegraphics[width=\textwidth]{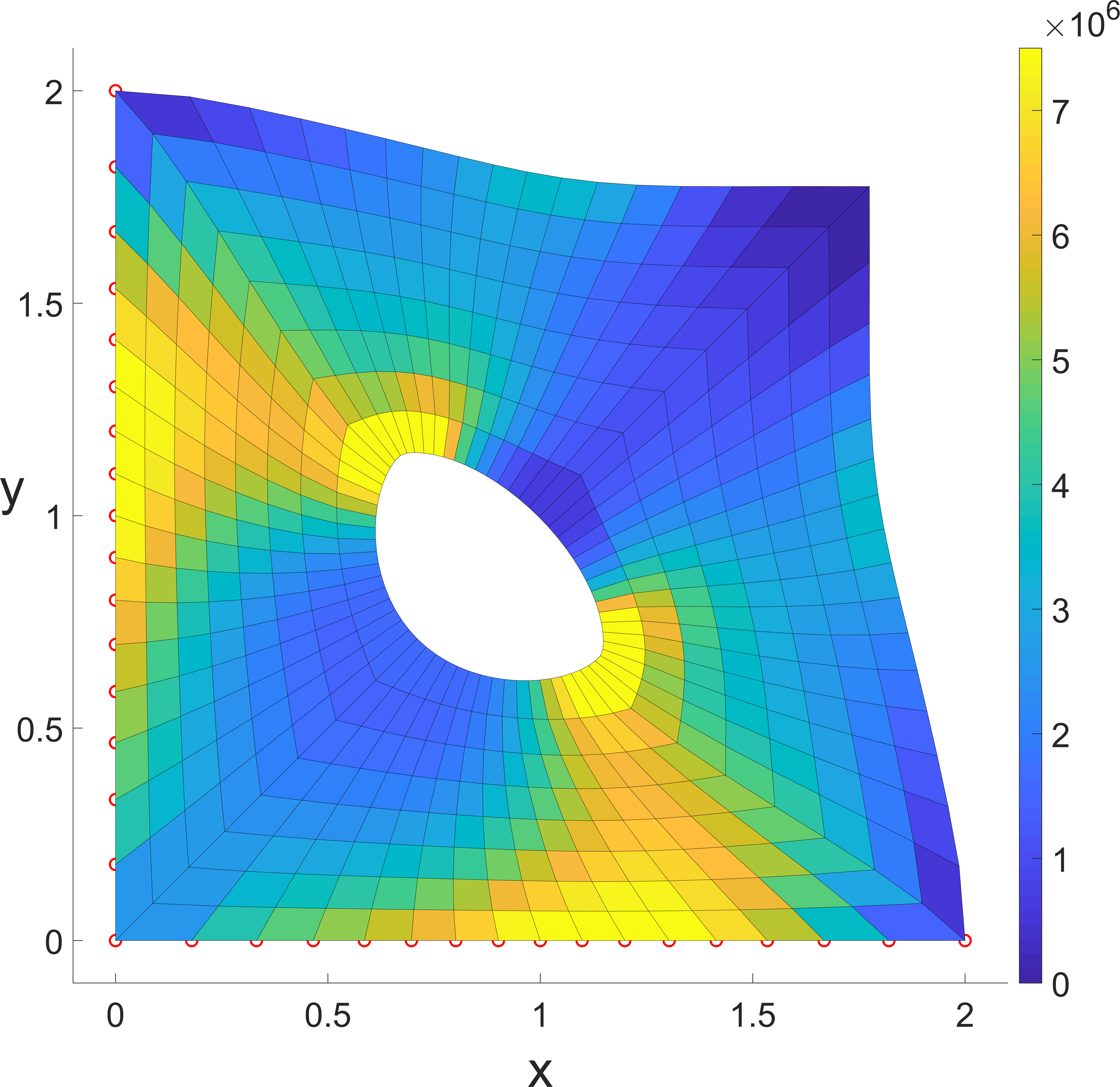}
\end{minipage} \:\:\:
\begin{minipage}[c]{0.445\textwidth}
\includegraphics[width=\textwidth]{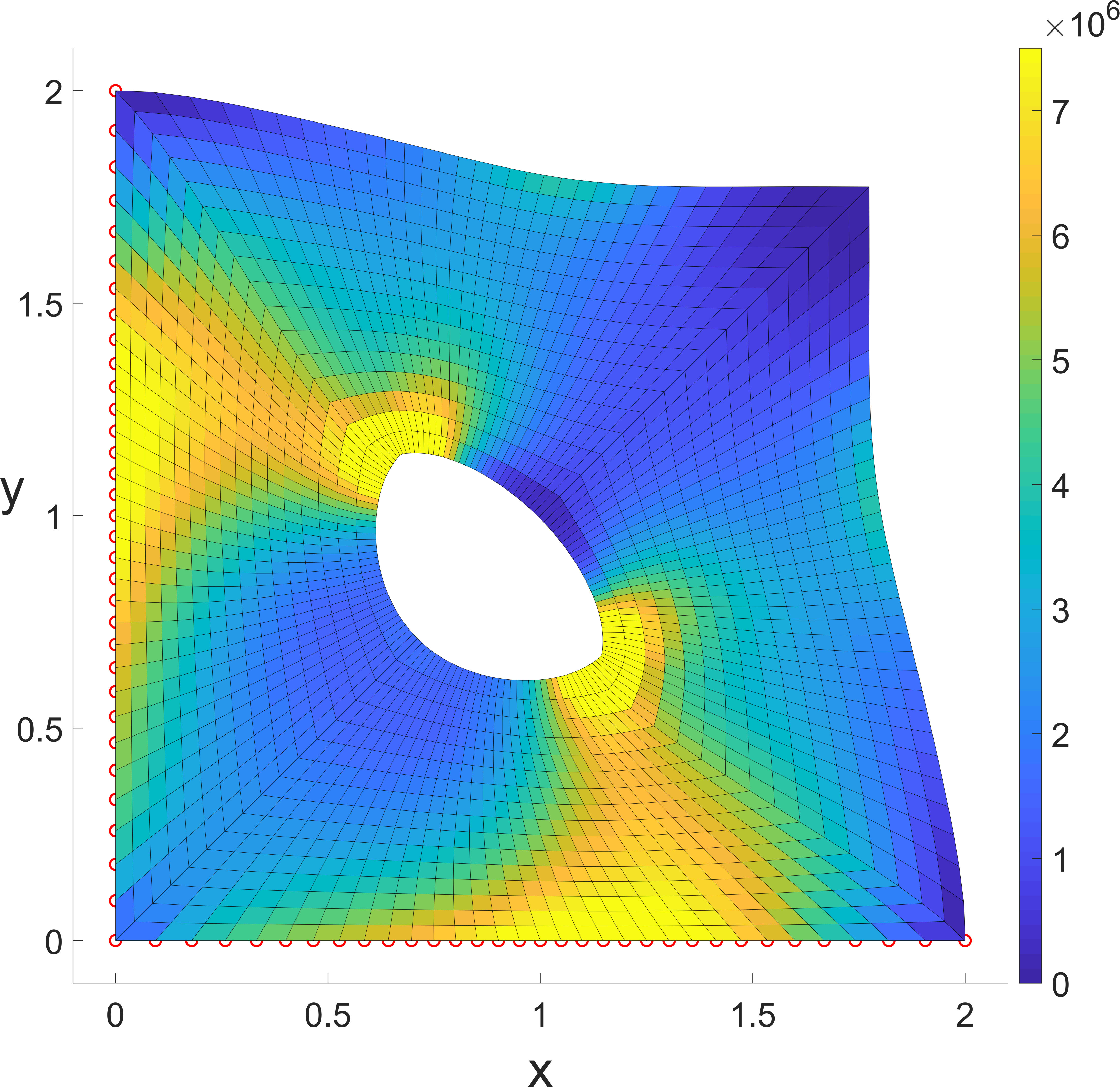}
\end{minipage}
\caption{Deformation and the corresponding Neo-Hookean density distributions for the 2D hyperelastic problem: 
a mesh with $\nt = 512$ and Q4 elements (left) and a mesh with $\nt = 2048$ and Q3 elements (right).}
\vspace{-0.4cm}
\label{fig:elasticity}
\end{figure}
\vspace{-3mm}
\noindent
\vspace{-3mm}

\begin{figure}[H]
    \centering
    \includegraphics[width=0.94\textwidth]{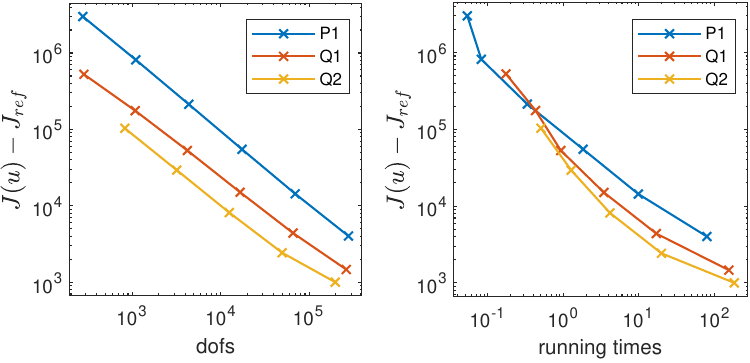}
    \caption{Performance of hyperelasticity: comparison of elements.}
    \label{fig:comparison_elasticity}
    \vspace{-0.2cm}
\end{figure}

We assume the same benchmark problem as in \cite{MoVa}: a 2D hyperelastic domain given by a square $[0,2] \times [0,2]$ perforated by a disk with radius $r = 1/3$ in its center is subjected to a constant volumetric vector force $\f = (-3.5 \cdot 10^7, -3.5 \cdot 10^7)$ acting in a bottom-left direction; zero Dirichlet boundary conditions are applied on the left and bottom edge. We assume the Young modulus $E = 2 \cdot 10^8$ and the Poisson ratio $\nu = 0.3$. We consider arbitrary, although mutually consistent physical units. For illustration, Fig. \ref{fig:elasticity} shows examples of the corresponding deformed mesh together with the underlying Neo-Hookean density distribution.
{\color{myrev}
Fig. \ref{fig:comparison_elasticity} depicts a comparison of P1, Q1 and Q2 elements. Similarly to Fig. \ref{fig:comparison_pLaplace}, Q2 elements are superior to Q1 and P1 in accuracy with respect to the number of dofs, however, we observe only a little improvement with respect to the evaluation times.}

\vspace{-2mm}

\subsection{Remarks on 2D implementation}
As an example, we introduce the following block that describes the evaluation of the p-Laplace energy:
\begin{listing}
\begin{lstlisting}
function e = energy(v,mesh,params)
v_elems = v(e2d_elems);      % values on elements in hp basis
F_elems = evaluate_F_scalar_2D(v_elems,Dphi_elems);  % grads
densities_elems = density_pLaplace_2D(F_elems,w,alpha);   
e = sum(areas_elems.*densities_elems) - b_full'*v;   % energy
end
\end{lstlisting}
\end{listing}

The structure corresponds exactly to the code of \cite{MoVa}. The main difference is that some objects inside \textbfn{evaluate\_F\_scalar\_2D} and \textbfn{density\_pLaplace\_2D} are higher-dimensional. This is because more Gauss points (we denote by $\nip$ their number) are needed for the integration of higher polynomial functions in the hierarchical basis.
The function \textbfn{evaluate\_F\_scalar\_2D} is provided below
\begin{listing}
\begin{lstlisting}
function F_elems = evaluate_F_scalar_2D(v_elems,Dphi)
v_elems3D = reshape(v_elems',size(v_elems,2),1,size(v_elems,1));
v_x_elems = Dphi{1}.*v_elems3D;  % np_ref x nip x ne
v_y_elems = Dphi{2}.*v_elems3D;  % np_ref x nip x ne
F_elems = cell(1,2);
F_elems{1,1} = squeeze(sum(v_x_elems,1));  % nip x ne
F_elems{1,2} = squeeze(sum(v_y_elems,1));  % nip x ne
end
\end{lstlisting}
\end{listing}
and evaluates the following objects:
\begin{itemize}
    \item (lines 3-4) 3D matrices \textbfn{v\_x\_elems} and \textbfn{v\_y\_elems} of size $\nbref \times \nip \times \nt$ store for every element the partial derivatives of local basis functions in all $\nip$ Gauss points multiplied by the corresponding values of $v$.
    \item (lines 6-7) the final matrices of size $\nip \times \nt$ storing the partial derivatives of $v$ in all Gauss points on every element.
\end{itemize}
The function \textbfn{density\_pLaplace\_2D} evaluating energy densities on every element is introduced below
\begin{listing}
\begin{lstlisting}
function densities = density_pLaplace_2D(F,w,alpha)
densities = (1/alpha)*sqrt(F{1,1}.^2 + F{1,2}.^2).^alpha; 
densities = densities'*w;
end
\end{lstlisting}
\end{listing}
and evaluates the following objects:
\begin{itemize}
    \item (line 2) the matrix of size $\nip \times \nt$ storing the p-Laplace energy densities in all Gauss points of every element.
    \item (lines 3) vector of length $\nt$ containing energy densities on all elements.
\end{itemize}
{\color{myrev}Similarly, the majority of the original P1 functions related to the evaluation of the hyperelastic energy are extended to the higher dimension in the same way}.
\vspace{-1mm}
\section{Conclusions and Outlook}
{\color{myrev}
The hp-FEM for 2D rectangular elements was successfully incorporated into our vectorized MATLAB code and its improved convergence performance was demonstrated on two examples. 

This work contributes to our long-term effort in developing a vectorized finite element-based solvers for energy minimization problems. Since many such problems emerge in science and engineering, the code is designed in a modular way so that various modifications (e.g., in functional types or boundary conditions) can be easily adopted. Our future research directions include implementing the hp-FEM in 3D or tuning the applied minimization algorithms.}

\bibliographystyle{abbrv}

\end{document}